# Characterization of a modular flow cell system for electrocatalytic experiments and comparison to a commercial RRDE system

Frederik J. Stender[a], Keisuke Obata[b], Max Baumung[a], Fatwa F. Abdi[b], Marcel Risch[a,c]*

[a] Frederik Johannes Stender, Max Baumung, Dr. Marcel Risch
Institut für Material Physik
Georg-August-Universität Göttingen
Friedrich-Hund-Platz 1, 37085 Göttingen

[b] Dr. Keisuke Obata, Dr. Fatwa Firdaus Abdi
Institut für Solare Brennstoffe
Helmholtz-Zentrum Berlin für Materialien und Energie GmbH
Hahn-Meitner-Platz 1, 14109 Berlin

[c] Dr. Marcel Risch
Nachwuchsgruppe Gestaltung des Sauerstoffentwicklungsmechanismus
Helmholtz-Zentrum Berlin für Materialien und Energie GmbH
Hahn-Meitner-Platz 1, 14109 Berlin
E-mail: marcel.risch@helmholtz-berlin.de

**Abstract:** Generator-collector experiments offer insights into the mechanisms of electrochemical reactions by correlating the product and generator currents. Most commonly, these experiments are performed using a rotating ring-disk electrode (RRDE). We developed a double electrode flow cell (DEFC) with exchangeable generator and detector electrodes where the electrode width equals the channel width. Commonalities and differences between the RRDE and DEFC are discussed based on analytical solutions, numerical simulations and measurements of the ferri-/ferrocyanide redox couple on Pt electrodes in a potassium chloride electrolyte. The analytical solutions agree with the measurements using electrode widths of 5 and 2 mm. Yet, we find an unexpected dependence on the exponent of the width so that wider electrodes cannot be analysed using the conventional analytical solution. In contrast, all the investigated electrodes show a collection efficiency of close to 35.4% above a minimum rotation speed or flow rate, where the narrowest electrode is most accurate at the cost of precision and the widest electrode the least accurate but most precise. Our DEFC with exchangeable electrodes is an attractive alternative to commercial RRDEs due to the flexibility to optimize the electrode materials and geometry for the desired reaction.

## Introduction

Generator-collector experiments offer insights into the mechanisms of electrochemical reactions by correlating the product and generator currents.[1,2] A common method is based on driving a redox reaction on the generator electrode and detection of the reverse reaction on the collector electrode[3–8], which works for fast redox couples[1,9] and detection of catalytic products.[10–13] Electrochemical detection has several requirements. First, the reaction needs to be chemically reversible in order to be detected. Second, possible side products or (semi)stable intermediate products must have a sufficiently different redox potential from the species in question to separate them.[5] Third, the transport of the product towards the collector should be well defined.

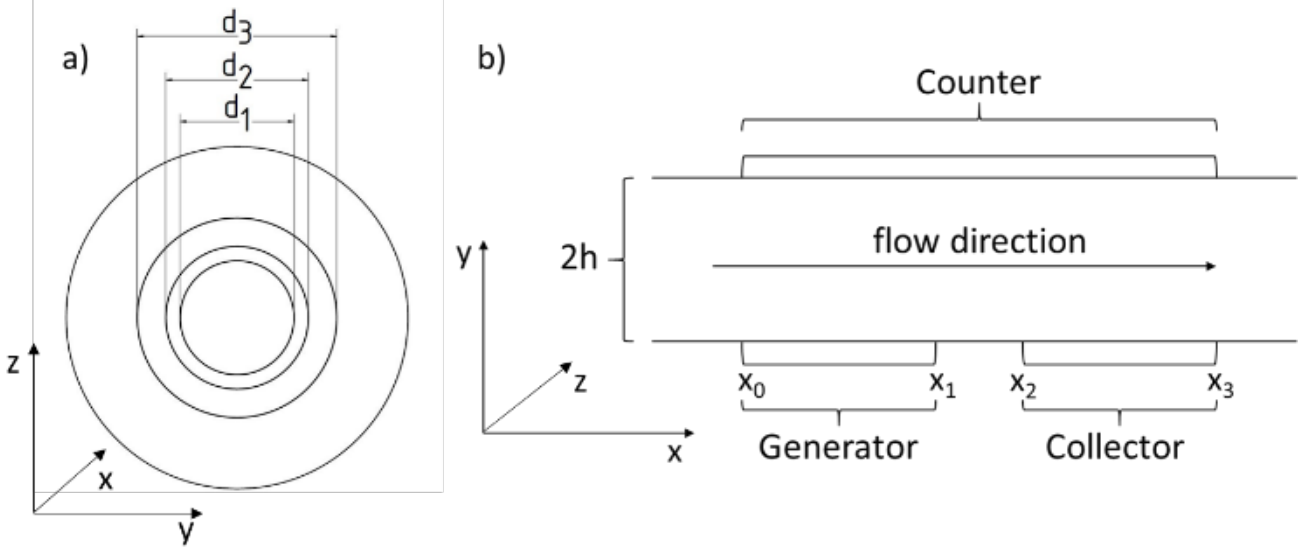

**Figure 1.** Schematic drawing of the used RRDE (a) and the flow cell (b). a) $d_1$ represents the diameter of the generator electrode (Disk), $d_2$ the diameter of the combination of generator and spacer and $d_3$ the diameter of the combination of the generator, spacer and collector (ring). b) $x_0$ represents the start of the generator electrode and $x_1$ the end. $x_2$ represents the start of the collector electrode and $x_3$ the end. The gap between both electrodes has the size $x_2$-$x_1$.

Generator-collector experiments are most commonly performed using rotating ring-disk electrodes (RRDE; Fig. 1a),[11–13] which are commercially available from several manufacturers. In this kind of setup, a rotated disk is the generator electrode. The rotation induces a laminar flow of the electrolyte on the disk. The collector electrode is a concentric ring, which is separated by a small isolating gap from the disk. The disk is exchangeable but not the ring.
An interesting but less popular setup for generator-collector experiments is the double electrode flow cell (DEFC),[1,14–17] where the generator and collector electrodes are embedded in a flow channel. Here, we only consider rectangular channels (Fig. 1b) but other geometries, e.g. tubes are also possible.[10,13,18] In well-constructed setups, a laminar flow transports the products from the generator to the collector electrode. Flow cells are also commercially available from several sources but only a single DEFC with two non-exchangeable disk electrodes is commercially available (for use in liquid chromatography). Self-made DEFCs offer the possibility to optimize the geometry for the target application and to make both the generator and collector electrodes exchangeable to study different reactions. Moreover, it is easy to integrate flow cells with additional analysis tools[19] such as UV-vis,[20,21] X-ray[22,23] or mass spectrometers[2,24] to gain more comprehensive insight into the electrochemical reactions at the generator electrode.
Here, we compare a commercial RRDE system with a self-made DEFC as an electroanalytic tool for generator-collector experiments. Our DEFC was designed in such a way that rectangular electrodes of different sizes can be easily clamped into and removed from the channel for post mortem analysis. The well-known ferro-/ferricyanide redox couple[25] was used as the analyte to compare both systems in terms of transport limiting currents and collection efficiency. Additionally, 3D computational simulations were conducted for the DEFC to get further insight into the flow behavior of the electrolyte, where the possible mixing of the analytes from the generator and counter electrodes was of special interest. We found that the proposed flow cell produces qualitatively similar results as the RRDE setup. We finally underlined a few significant differences in the hydrodynamic behavior that one needs to be aware of when comparing or selecting both systems.

## Results and Discussion

### Theory

The transport of the analytes within the cell needs to be understood to predict and understand the measured currents. The DEFC and RRDE systems are both well-defined hydrodynamic systems, in which diffusion as well as convection determine the transport of the analytes towards the electrodes.
The faradaic current of a fast redox reaction (i.e. without coupled kinetics)

$$\boldsymbol{A} + ne \rightarrow \boldsymbol{B} \qquad (1)$$

is given by

$$i = nF \int J \, dS \qquad (2)$$

where $i$ is the current, $n$ is the number of electrons involved in the redox-reaction, $F$ is the Faraday constant, $S$ is the area and $J$ is the flux of the redox couple towards the electrode.
The nature of convection differs among the RRDE and DEFC so that the achievable limiting currents are affected differently by forced convection. Laminar flow is assumed in the theoretical treatment of both systems in textbooks[1,26,27] as presented in the following first for the RRDE and then for the DEFC.

### Analytical solutions of the RRDE

The laminarity of the RRDE can be approximated using the Reynolds number

$$Re_{RRDE} = \frac{\omega r^2}{\upsilon}, \qquad (3)$$

where $\omega$ is the rotation speed in units of Hz, $r$ is the radius of the disk, and $\upsilon$ is the kinematic viscosity of the solution. The Reynolds number should not exceed 2000 to avoid turbulent flow.[28]
The convection-diffusion equation describes the concentration change of species A (Eq. 1) due to convection and diffusion. It is given by

$$\frac{\partial[\boldsymbol{A}]}{\partial t} = D_{\boldsymbol{A}} \nabla^2 [\boldsymbol{A}] - \left( v_x \frac{\partial[\boldsymbol{A}]}{\partial x} + v_y \frac{\partial[\boldsymbol{A}]}{\partial y} + v_z \frac{\partial[\boldsymbol{A}]}{\partial z} \right), \qquad (4)$$

where $D_{\boldsymbol{A}}$ is the diffusion constant of species **A**, $[\boldsymbol{A}]$ is the concentration, $v_{x,y,z}$ are the velocity profiles in the $x, y, z$ directions as defined in Fig. 1a. Under steady state conditions, the time dependency is removed as there is no change in the velocity and concentration profiles over time.
For the RRDE system, only the x-direction contributes to the transport of materials toward the electrode, assuming the electrode is uniformly accessible, which simplifies the convection-diffusion equation, i.e. Eq. (4), to

$$0 = D_A \frac{\partial^2 [A]}{\partial x^2} - v_x \frac{\partial [A]}{\partial x}. \qquad (5)$$

The velocity profile in $x$-direction close to the electrode can be approximated by[28,29]

$$v_x \approx -0.510(2\pi\omega)^{3/2}\nu^{-1/2}x^2. \tag{6}$$

With the mass transport limiting boundary condition ([$A$] = 0) at the electrode surface, the solution of Eq. 2 and Eq. 5 using the approximation in Eq. 6 results in the limiting current equation for a first-order reaction on the disk of the RRDE[28]

$$|i_{\mathrm{lim,RRDE}}| = \alpha nFAD^{\frac{2}{3}}\nu^{-\frac{1}{6}}\omega^{\frac{1}{2}}c_\infty. \tag{7}$$

Where $\alpha$ is a pre factor that comes from the geometry and flow velocity distribution and has the value 1.554 ($\omega$ in Hz), 0.620 ($\omega$ in angular velocity) or 0.201 ($\omega$ in rpm)[30]. $A$ and $c_\infty$ are the electrode area and concentration in the bulk solution, respectively. The corresponding diffusion layer thickness ($z_D$) is constant due to the uniform accessibility of the electrode. It is given by

$$z_D = \beta D^{1/3}\nu^{1/6}\omega^{-1/2}, \tag{8}$$

where $\beta$ has the value 0.643 ($\omega$ in Hz), 1.613 ($\omega$ in angular velocity) or 4.975 ($\omega$ in rpm) The collection efficiency describes the ratio of the collector current (ring of an RRDE) to the generator current (disk of an RRDE). In a chemically and electrochemically reversible redox system, the experimental collection efficiency is thus given by

$$N_{\mathrm{exp}} = -\frac{i_{\mathrm{collector}}}{i_{\mathrm{generator}}}. \tag{9}$$

The theoretical collection efficiency of the RRDE system, $N_{\mathrm{RRDE}}$, is calculated analytically from three diameters: the outer disk diameter $d_1$, the outer gap or inner ring diameter $d_2$ and the outer ring diameter $d_3$ (as illustrated in Fig. 1a). It is given by[10]

$$N_{\mathrm{RRDE}} = 1 - \sigma^2{}_{OD} + \sigma^{\frac{2}{3}}{}_B - G(\sigma_C) - \sigma^{\frac{2}{3}}{}_B G(\sigma_A) + \sigma^2{}_{OD} G(\sigma_C \sigma^3{}_{OD}), \tag{10}$$

with the function

$$G(x) = \frac{1}{4} + \left(\frac{\sqrt{3}}{4\pi}\right) ln\left[\frac{\left(x^{\left(\frac{1}{3}\right)}+1\right)^3}{x+1}\right] + \left(\frac{3}{2\pi}\right) arctan\left[\frac{2x^{\frac{1}{3}}-1}{\sqrt{3}}\right], \tag{11}$$

where

$$\sigma_{OD} = \frac{d_3}{d_1}, \sigma_{ID} = \frac{d_2}{d_1}$$

$$\sigma_A = \sigma^3{}_{ID} - 1, \sigma_B = \sigma^3{}_{OD} - \sigma^3{}_{ID}, \sigma_C = \frac{\sigma_A}{\sigma_B}.$$

Common commercial RRDEs with interchangeable disks have dimensions of ($d_1$ = 5.0 mm, $d_2$ = 6.5 mm, $d_3$ = 7.5 mm) resulting in a collection efficiency of N = 25.6 % and ($d_1$ = 4 mm, $d_2$ = 5 mm, $d_3$ = 7 mm) resulting in a collection efficiency of 42 %. The latter commercial RRDE (Fig. 1a) is used in this study.

**Analytical solutions of the DEFC**

The derivation for a rectangular DEFC follows the same steps but different boundary conditions apply. For the DEFC, the Reynolds number is expressed as:

$$Re_{DEFC} = \frac{2hv_0}{\nu} = \frac{3V_f}{2\nu d}, \tag{12}$$

where $h$ is the half-height of the channel, $v_0$ is the velocity at the center of the channel (y=h, z=d/2), $\nu$ is the kinematic viscosity of the solution, $d$ is the width of the channel and $V_f$ is the volume flow rate

For the DEFC, the diffusion in $x, z$-direction can be neglected due to fast convection in comparison to diffusion[26,28,31]. Additionally, it is assumed that the velocity in all other directions is negligible ($v_y$=$v_z$=0) for laminar flow in $x$ direction. This reduces Eq. (4) to

$$0 = D_A \frac{\partial^2[A]}{\partial y^2} - v_x \frac{\partial[A]}{\partial x}. \tag{13}$$

The parabolic velocity profile in a rectangular channel can be linearly approximated close to the electrode using the Leveque approximation,[1,28,32,33] i.e.,

$$v_x = v_0\left(\frac{h^2 - (y-h)^2}{h^2}\right) \approx \frac{2v_0 y}{h} = \frac{3V_f y}{2h^2 d} \tag{14}$$

for $y$ = 0 with

$$V_f = \frac{4}{3}v_0 hd. \tag{15}$$

Eq. 14 is plotted in (Fig. 2a) for illustration. The solution of Eqs. 2 and (13) using the approximation in Eq. (14) results in the equation that describes the limiting current of a first-order reaction[28]

$$|i_{\mathrm{lim,DEFC}}| = \alpha n F w D^{\frac{2}{3}}\left(\frac{V_f}{h^2 d}\right)^{\frac{1}{3}} l^{\frac{2}{3}} c_{\infty}, \tag{16}$$

where $w$ is the width of the active electrode area, $c_{\infty}$ is the bulk concentration, $l$ is the length of the electrode and $\alpha$ is a prefactor that comes from the geometry and flow velocity distribution and has the value $\alpha$=0.925 (h being the half height of the channel). A prefactor of $\alpha$=1.467 can be found in literature when h is the height of the channel. The corresponding diffusion layer thickness at the electrode can be expressed as[31]

$$z_D = \beta h \left(\frac{xDd}{hV_f}\right)^{1/3}. \tag{17}$$

Where the prefactor $\beta$ has the value 1.62162.

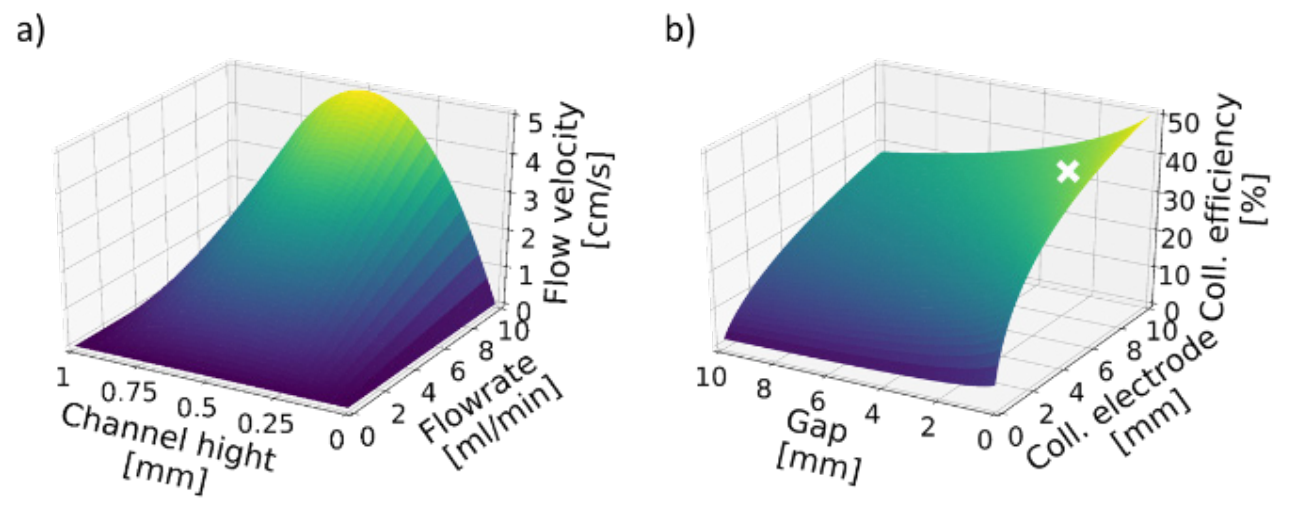

**Figure 2.** a) Analytically calculated flow field distribution over the height of the channel for different flow rates. b) Expected collection efficiency as function of the collection electrode and gap size according to analytical calculations as detailed in the text (Eq. 18). The white cross marks the collection efficiency of the experimentally used flow cell in this study.

The theoretical collection efficiency of a DEFC, $N_{\mathrm{DEFC}}$, for electrochemically reversible reactions can be analytically calculated as[10]

$$N_{\mathrm{DEFC}} = 1 + \lambda^{\frac{2}{3}}[1 - F(\theta)] - (1 + \theta + \lambda)^{\frac{2}{3}} \cdot \left[1 - F\left(\left(\frac{\theta}{\lambda}\right)(1 + \theta + \lambda)\right)\right] - F\left(\frac{\theta}{\lambda}\right), \tag{18}$$

with

$$\theta = \frac{x_2 - x_1}{x_1}$$
$$\lambda = \frac{x_3 - x_2}{x_1}$$

and the function

$$F(\theta) = \frac{\sqrt{3}}{4\pi} \ln\left[\frac{\left(1 + \theta^{\frac{2}{3}}\right)^3}{1 + \theta}\right] + \frac{3}{2\pi} tan^{-1}\left[\frac{2\theta^{\frac{1}{3}} - 1}{\sqrt{3}}\right] + \frac{1}{4}, \tag{19}$$

where $x_1$ is the length of the generator electrode, $x_2$ is the sum of the length of the generator electrode and the gap length and $x_3$ is the sum of the length of the generator electrode, the gap length and the length of the collection electrode (Fig. 1b). The analytically calculated collection efficiencies for varying gap lengths (effects parameter $x_2$, $x_3$, between 0.1 and 5 mm) and lengths of the detection electrode (effects parameter $x_3$, from 0.1 to 10 mm) are plotted in Fig. 2b. The resulting collection efficiencies range from 2.3% (5 mm gap, 0.1 mm detection electrode) to 50% (0.1 mm gap, 10 mm detection electrode).

The theory for the RRDE and DEFC systems show different dependencies. The limiting current of the RRDE explicitly depends on viscosity due to the velocity distribution in the electrolyte but not the geometry (except surface area). In contrast, the limiting current of the DEFC depends on the geometry but not on the viscosity, which only determines the pressure drop in the DEFC channel at any given flow velocity.

**Construction of the DEFC**

The experimental setup of the developed DEFC implements a four-electrode setup along a channel with two 5 mm long working electrodes on the floor of the channel separated by a 1 mm gap. The upstream working electrode serves as the generator electrode and the downstream working electrode as the collector electrode. An 11 mm long counter electrode is situated directly above the two working electrodes (Fig. 1b). The geometrical parameters of our flow cell are summarized in Table 1. The working electrodes are clamped into a modular body and tightened with PTFE-Tape and Parafilm® as gaskets. The clamping allows easy removal of the electrodes for post-mortem investigations, and a wide range of differently prepared electrodes can be used without modification of the cell body. As the electrodes are clamped, their width matches that of the channel, which has implications that are discussed below. Electrode widths of 2.0 mm, 5.0 mm and 7.5 mm were selected for this study. Yet, the physical size remained identical so that a smaller width results in a smaller utilized fraction of the electrode, i.e. active area. The reference electrode was placed upstream and close to the generator electrode to avoid issues with product interference (if gas would be produced) and to accurately reflect the potential at the generator electrode. Moreover, the geometry of parallel working and counter electrodes was chosen to ensure homogeneous equipotential surfaces parallel to the electrode surfaces, which minimize the error of the potential measurement and ensure better stability. For flow rates of 0.5 to 8 ml/min, the Reynolds number is calculated as $Re_{DEFC}$ = 2.5-40 ($\upsilon = 0.01004\ \mathrm{cm}^2/s$ at 20 °C) using Eq. (12). The low Reynolds number (<<2000) suggests that the cell has a laminar flow, which allows a controlled mass transport to and from the working electrodes.

**Table 1**. Geometric parameters of the RRDE and DEFC

| | **RRDE** | | **DEFC** | |
|---|---|---|---|---|
| **Description** | **Variable** | **Value** | **Variable** | **Value** |
| Length of generator electrode | $d_1/2$ | 4 mm | $x_1 - x_2$ | 5 mm |
| Width of generator electrode | $d_1/2$ | 4 mm | $z_1 – z_0$ | 2.0, 5.0*, 7.5 mm |
| Gap | $d_2$-$d_1$ | 1 mm | $x_0 – x_1$ | 1 mm |
| Length of collector electrode | ($d_3$-$d_2$ )/2 | 2 mm | $x_3 – x_2$ | 5 mm |
| Width of generator electrode | ($d_3$-$d_2$)/2 | 2 mm | $z_3 - z_2$ | 2.0, 5.0*, 7.5 mm |
| Channel height | n/a | n/a | 2h | 1 mm |

*Used channel width for the experiments.

**Numerical simulations of the DEFC**

Multiphysics finite-element simulations were performed to initially check the applicability of the analytical calculations above and to secondly gain further insight into the mass transport in the DEFC. Identical parameters are used in the calculations and simulations (Table 2). The simulations do not require some of the assumptions made in the analytical calculations as discussed below regarding laminar flow, the velocity profiles as well as the current and product distributions.

**Table 2.** Parameters of analytical calculations and simulations.

| Description | Variable | Value |
|---|---|---|
| Bulk concentration $[Fe(CN_3)_6]^{3-}$ | $c_{\infty,[F_e(CN_3)_6]^{3-}}$ | 2 mM |
| Diffusion coefficient $[Fe(CN_3)_6]^{3-}$ | $D_{[F_e(CN_3)_6]^{3-}}$ | 7.31E-6 $cm^2/s$ |
| Diffusion coefficient $[Fe(CN_3)_6]^{4-}$ | $D_{[F_e(CN_3)_6]^{4-}}$ | 6.71E-6 $cm^2/s$ |
| Diffusion coefficient $O_2$ | $D_{O_2}$ | 2.40E-5 $cm^2/s$ |
| Kinematic viscosity of water | $\nu$ | 0.010023 $cm^2/s$ |

The analytical calculations predicted laminar flow due to the low Reynolds number. Yet, the liquid inlet may cause a disturbance to the laminar flow. Simulations of our cell geometry show that the disturbance of the inlet only affects regions within ~5 mm from the inlet (Fig. S1). As the inlet is separated by 35 mm from the generator electrode, we conclude that the flow is uniform and laminar over the 5 mm length (x dimension in Fig. 1b) of the generator and also collector electrode. The 2D analytical calculations assume uniform velocity in the direction of the channel width ($z$ direction in Fig. 1b) but the velocity was reduced in the simulations near the walls (z=0.0 and z=5.0 mm). This is shown in Fig. 3a for half the channel height (y coordinate) and half the electrode width (z coordinate). The simulation is symmetric relative to these half points.

For the shown simulation of a 5.0 mm wide electrode, the wall affects only the leftmost 0.5 mm (dashed white line), i.e., about 20% of the electrode width. Thus, the assumption of a uniform velocity in the direction of the channel width is reasonably fulfilled for the 5 mm and potentially larger electrodes.

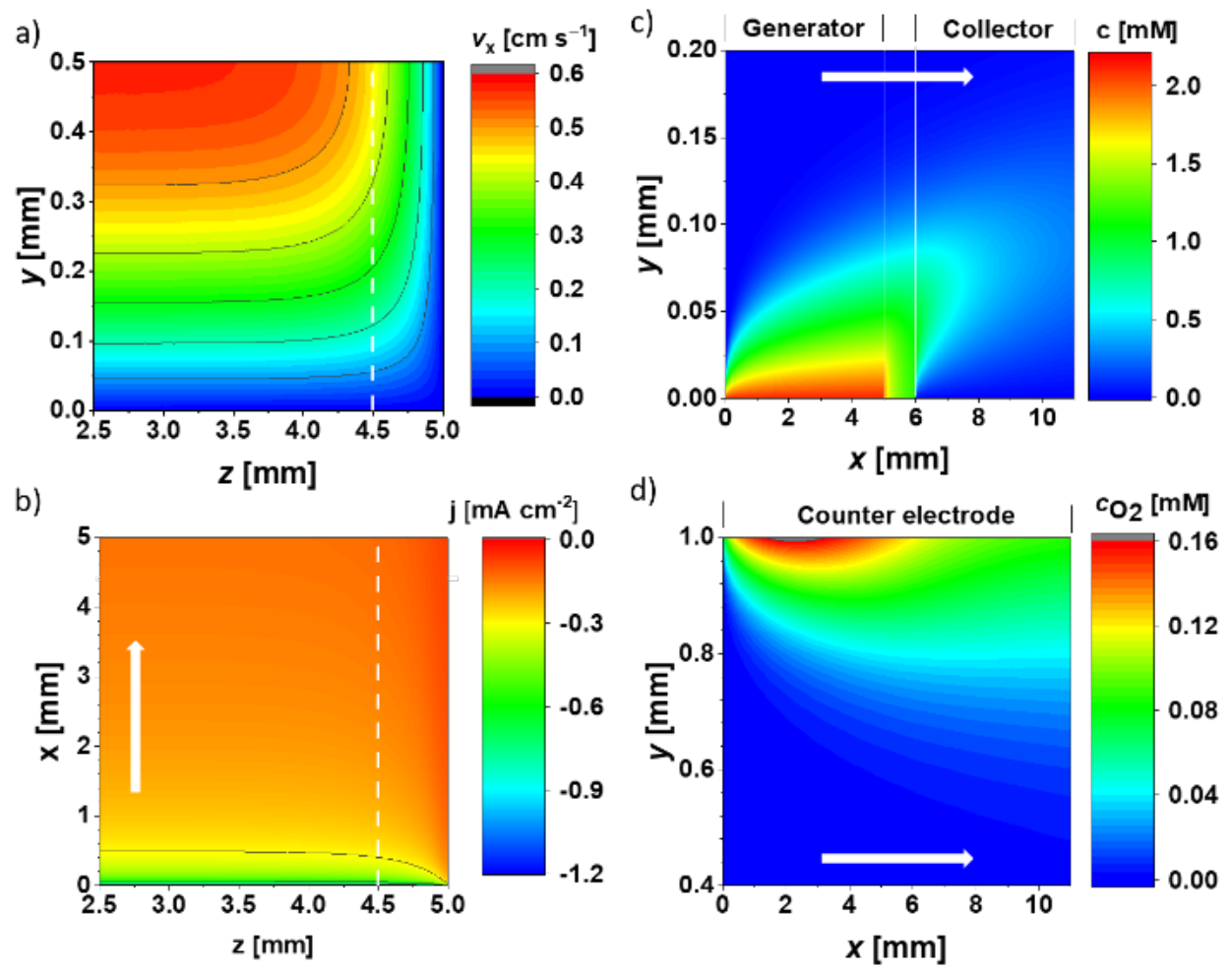

**Figure 3.** Numerical simulations of the DEFC (w = 5mm). a) Flow velocity field at a flow rate of 1 ml/min along the width (z) and the height of the channel (y) due to the symmetry of the system only the half width and height are displayed. b) Local current density on the generator at 1 mL/min. The symmetry boundary is located at $z$ = 2.5 mm. c) Concentration distribution of $Fe(CN)_6^{4-}$ in the middle of the channel ($z$ = 2.5 mm) at 1 mL/min d) Concentration distribution of dissolved $O_2$ at 1 mL/min at the middle ($z$ = 2.5 mm) of channel. The white arrows symbolize the flow direction (x-direction). The vertical line in panel a and b indicates the region with the most pronounced changes due to the wall.

The current density distribution over the generator electrode resulting from the non-uniform velocity profile is shown in Fig. 3b, where again the symmetric half point is shown for the electrode width (z dimension) and length (x dimension / flow direction). The current density decreases from -1.2 mA/cm$^2$ at the electrode edge toward the inlet to -0.15 mA/cm$^2$ within about 2 mm in flow direction, where the vast majority of the current occurs in the first 0.5 mm. The current density is reduced toward the wall (z=5.0 mm) but since the reduction is constraint to the first 10% in flow direction, the walls likely have only a minor effect on the current density when the electrode is long and wide as in our design.
The product distribution at both working electrodes (Fig. 3c) was subsequently simulated. At the generator electrode, ferricyanide ($Fe^{III}[CN]_6^{3-}$) is reduced to ferrocyanide ($Fe^{II}[CN]_6^{4-}$) and the reverse reaction occurs at the collector electrode. The concentration of ferrocyanide is highest near the generator electrode (y=0.0 mm) and decays to the bulk value of zero within a height of 0.15 mm above the electrode (y dimension). In the direction of the flow (x dimension), the concentration increased parabolic in height (y). The ferrocyanide concentration was about 1 mM in the gap to a height of 0.07 mm and then quickly decayed to zero. The concentration of ferrocyanide also vanished near the collector electrode as expected. The contour line of 0.5 mM (light teal) illustrates that the concentration decrease is again parabolic in height. Finally, we simulated the product distribution above the counter electrode located at y=1.0 mm (Fig. 3d). An oxidation reaction must occur for charge conservation at the counter electrode because a reduction reaction occurs at the generator electrode and the collection efficiency of the collector electrode is <100%. Since ferricyanide cannot be oxidized further, the most likely component in the electrolyte to oxidize is water by the oxygen evolution reaction ($2H_2O \rightarrow 4H^+ + 4e^- + O_2$). For low and intermediate $O_2$ concentrations (< 0.04 mM), the distribution is again nearly parabolic in height (y dimension). The highest concentrations of 0.12 to 0.16 mM were observed above the generator electrode (0<x<5 mm) with a semicircular distribution. Overall, the simulations of the product distributions show that the products of the working and counter electrodes do not mix for the chosen channel height of 1 mm.

### Comparison of DEFC and RRDE

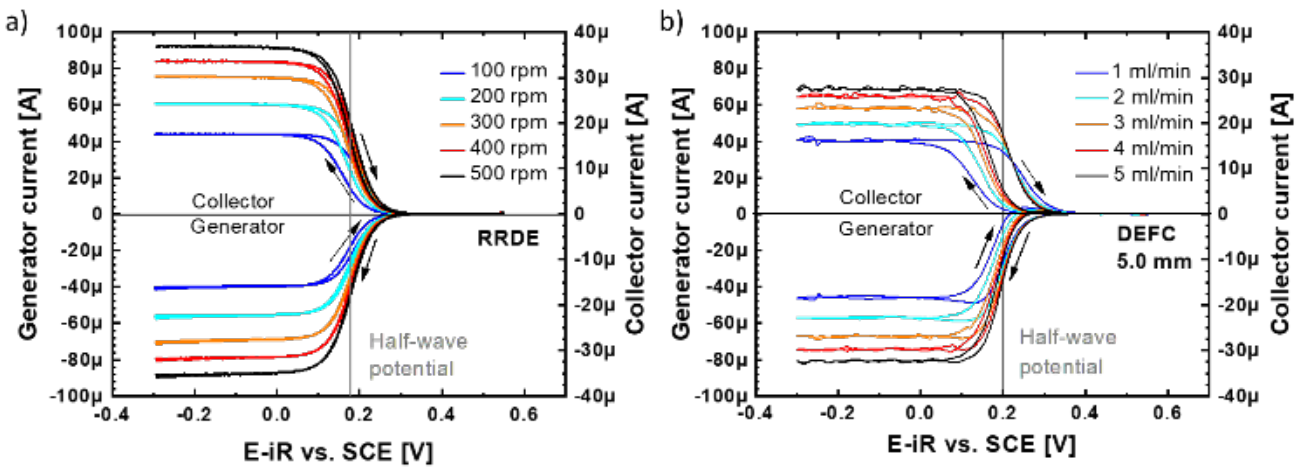

**Figure 4.** Comparison of the flow cell to an RRDE system. (a) RRDE measurements with rotation speeds between 100 and 500 rpm. (b) Flow cell measurements with flow rates between 1-5 ml/min.

We assembled the DEFC with generator and collector electrodes made from Pt to test its response with ferricyanide in a potassium chloride supporting electrolyte. Likewise, CV measurements on a RRDE were performed with a Pt disk (generator) and ring (collector) in the same electrolyte using rotation speeds that result in comparable limiting currents for the generator and collector electrodes in both the RRDE (Fig. 4a) and DEFC with w=5.0 mm (Fig. 4b). The measurements of the DEFC with widths 2.0 and 7.5 mm are shown in Fig. S2. The DEFC measurements are precisely reproduced on different laboratory days (Fig. S3). All measurements show exponential currents at the generator and collector electrodes with opposite signs between 0.3 and 0.1 V vs. a saturated calomel electrode (SCE), where mainly the kinetics of the ferri-/ferrocyanide redox limit the currents. The currents in both systems were constant below 0.1 V vs. SCE due to limited mass transport of the ferricyanide analyte.
The half-wave potentials at both the DEFC and RRDE decrease slightly by ~10 mV with increasing flow rate and rotation speed, which suggests partial limitation by mass transport at the investigated low flow rates and rotation speeds. The RRDE gives a value of 0.18 V vs. SCE for the half-wave potential at 200 rpm. For comparable currents at 2 ml/min the DEFC gives a slightly higher value of 0.2 V vs SCE. Thus, they were similar to the literature value[34] of 0.19 V.
A closer inspection of the data shows that the DEFC produced noisier data and exhibits stronger hysteresis as compared to the RRDE. The source of the noise are small fluctuations of the flow velocity due to the used syringe pump and slipping of the syringe plunger. The noise is visible but minor and does not influence the analyses below, which are based on average currents to reject the noise. The hysteresis was more pronounced for slower flow rates and smaller electrode width (Fig. S2, 4b). Its presence may seem surprising as the generator and collector electrodes are separated by 1 mm in both systems, yet the relevant flow velocity differs among the DEFC and RRDE so that the time of travel is longer for a DEFC at a comparable limiting current. As can be seen at the velocity distribution close to the electrode surface (Fig. S5). Furthermore, the DEFC electrodes are longer in flow direction and have a non-uniform current distribution (Fig. 3b). The hysteresis also does not affect the limiting current at the working electrodes and the corresponding collection efficiency that are discussed in detail below.

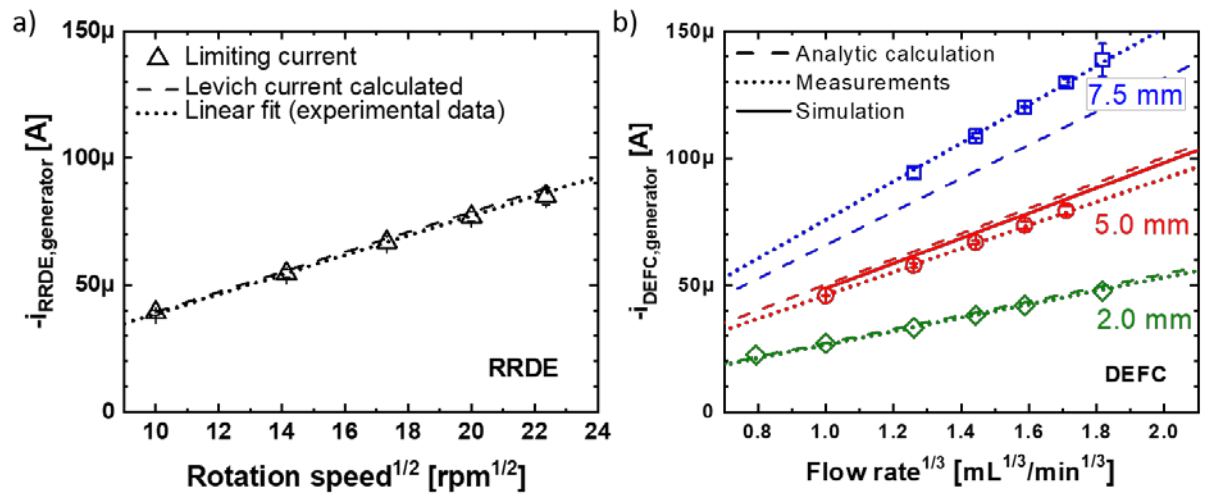

**Figure 5** a) Analytically calculated and measured RRDE limiting currents for different rotation speeds. b) Comparison between analytically calculated, simulated and measured limiting currents for different flow rates and channel widths. The plots are linearized by using the expected exponential function of the flow rate or rotation speed (Eq. 7,16).

**Limiting current of the DEFC and RRDE**

The measured limiting current of the generator electrodes for the RRDE and DEFC was compared to the analytical solution as well as simulations, which used typical parameters from literature (Table 2). As expected from Eq. 7 and 16, the absolute value of the limiting currents of the RRDE (Fig. 5a) and DEFC (Fig. 5b) increased with rotation speed and flow velocity, respectively. The limiting generator currents in both systems had similar values for the displayed flow rates and rotation speeds (Fig. 5). However, the active area of the electrode was 0.126 $cm^2$ for the RRDE system but 0.25 $cm^2$ for the flow cell. This indicates, that the diffusion layer must be notably smaller in the RRDE system at these rotation speeds. This is also expected from the theory of the diffusion-layer thickness (Eq. 8 and 17). A thickness between 20 µm (500 rpm) and 45 µm (100 rpm) is expected (Eq. 8) for the RRDE while 49 µm (5 mL/min) to 84 µm (1 mL/min) were calculated (Eq. 17) at the middle of the 5 mm DEFC.
The slopes in Fig. 5 relate to physical constants of the electrolyte and the geometry of the setup as given by Eq. 7 and 16 in the analytical treatment. Table 3 compiles all obtained values. The experimental fit assumed that the fit passes through the origin like the analytical equations. Alternative fits with finite y-axis intercept may be found in Table S1 and Fig. S4. As the $R^2$ values of both models were close to unity, indicating an excellent fit, we discuss the simpler model corresponding to the analytical formulas. For the RRDE, the measured and analytical slopes of the generator electrode agree well (Fig. 5a), which indicates that the physical constants can be determined with reasonable accuracy using Eq. 7. For the DEFC, the agreement between the measurement and the analytical calculation depended on the width of the generator electrode, where the narrowest electrode (w=2.0 mm) was described best, while the slope was slightly overestimated for w=5.0 mm and clearly underestimated for w=7.5 mm (Fig. 5b). The simulation of the DEFC with w =5.0 mm resulted in a slope closer to the analytical solution than the measurement. We thus focus on the analytical solutions. The treatment of the generator electrode of the DEFC necessitates to revisiting Eq. 17. Since the channel width (*d*) and electrode width (*w*) of our DEFC are identical, the analytical equation becomes

$$|i_{\text{lim,DEFC}}| = 0.925F(nc_\infty)^1(wDl)^{2/3}(V_f/h^2)^{1/3}. \quad (20)$$

where the variables are defined in the section "analytical solutions of the DEFC". The variables with the highest exponent and thus largest contribution to Eq. 20 are $n$, and $c_\infty$(power 1) and $w$, $D$, $l$ (power 2/3). We do not expect any significant deviations of the ferricyanide concentration (≤2%) and the length of the electrode (≤3%). We cannot separate the influences of deviations in the number of transferred electrons ($n$) and the diffusion constant ($D$). However, these parameters would have the same effect on the RRDE measurements in the same solution, where the slope agreed well with the analytical solution (about 2.5% deviation). This leaves the width of the electrode and the design decision of a channel width equal to the active electrode area as the most likely source of the deviation.

**Table 3.** Slopes of current against rotation speed or flow rate obtained from the data in Fig. 5.

| | **Analytic*** | **Simulation** | **Experimental fit y = ax+0** | |
|---|---|---|---|---|
| **Device** | **slope** | **slope** | **slope a** | **$R^2$** |
| RRDE | -3.96 μA/rpm$^{1/2}$ | n/a | -3.86(2) μA/rpm$^{1/2}$ | 0.9999 |
| DEFC, w = 2.0 mm | -27.3 μA min$^{1/3}$/cm | n/a | -26.6(2) μA min$^{1/3}$/cm | 0.9996 |
| DEFC, w = 5.0 mm | -50.2 μA min$^{1/3}$/cm | -49.5 μA min$^{1/3}$/cm | -46.0(1) μA min$^{1/3}$/cm | 1.0000 |
| DEFC, w = 7.5 mm | -65.8 μA min$^{1/3}$/cm | n/a | -75.7(0) μA min$^{1/3}$/cm | 1.0000 |

The determined experimental slopes are not proportional to $w^{2/3}$ (Fig. S6) as expected from Eq. 20. The deviation to the analytical solution was larger for wider electrodes. The solution of the limiting current in the limit of a large electrode is[10]

$$|i_{\text{lim,DEFC}}^{large}| = F(nc_\infty)^1 \mathrm{V}_\mathrm{f}^1, \quad (21)$$

which does not depend on $w$, i.e., the exponent of w decreases from 2/3 to 0, while the exponent of the flow velocity increases from 1/3 to 1 in this limit. As $w^{2/3}$ is < 1 (Fig. S6), a decrease of this exponent is expected to lead to a larger slope as observed. Thus, we conclude that the assumptions of the analytical calculation are unsuitable to describe the limiting current of the 7.5 mm DEFC, while it can be used for the narrower electrodes with w = 5.0 and 2.0 mm.
The changing dependence on the electrode width has important implications for the normalization. It is good practice to normalize the RRDE currents by the electrode area, $A$, which is the only geometric variable in Eq. 7. In contrast, the dependence of the diffusion layer thickness on the geometry is more complicated for the DEFC and the width, length and height contribute differently to the limiting current (Fig. 3b). The dependence on geometry in the limits of the analytical solution can be seen generally in Eq. 16 and specifically for our setup in Eq. 20. Normalization of the measured DEFC currents by the geometry factor

$$f = \frac{(wl)^{2/3}}{h^{2/3}} \quad (22)$$

gives reasonable results for w=2.0 mm and w=5.0 mm but clearly deviates for w=7.5 mm (Fig. S7) due to a different dependence on the width, w. Thus, the desirable normalization by a geometric factor is only meaningful for DEFCs if the exact dependence on the geometric variables is known.

**Collection efficiency of the DEFC and RRDE**

The measured limiting current of the collector electrodes and the corresponding collection efficiency for the RRDE and DEFC were compared to the analytical calculations as well as simulations (Fig. 6 and S8). The collection efficiency (open symbols in Figs. 6a and S8a) of the RRDE was identical within error to the analytical value of 42.5% for rotation rates >250 rpm (16 rpm$^{1/2}$) and slightly higher for slower rotation. The collection efficiency was also determined using Eq. 10 (Fig. S9a, Table 4). Allowing a finite y-axis intercept did not improve the fits (Table S2). The experimental collection efficiency of the RRDE matched the analytical calculation within 3σ. The collector current was calculated using the analytical collection efficiency and the measured generator currents. This prediction (dashed line) consequently matched the measured collector currents (solid symbols) well.

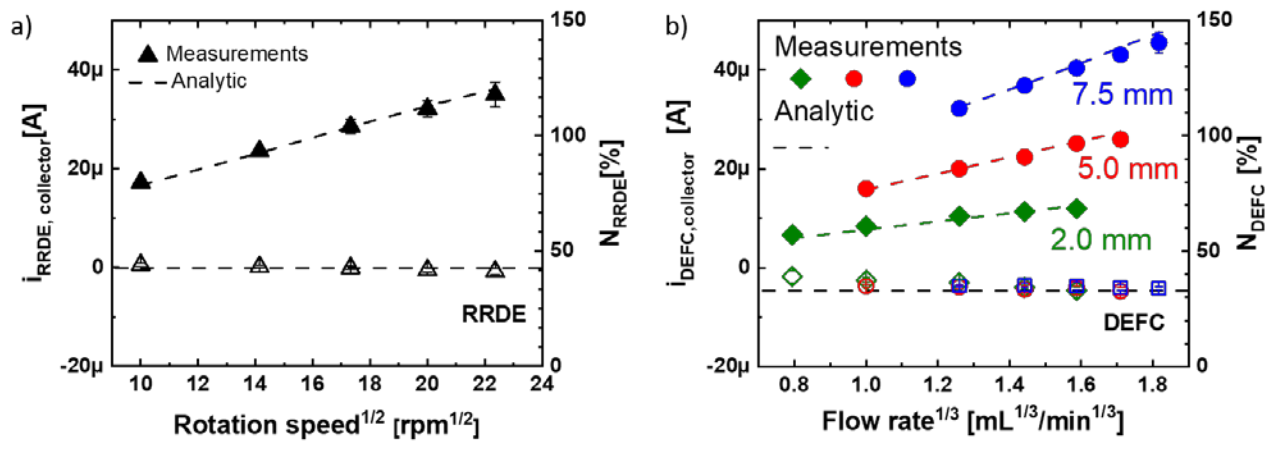

**Figure 6.** Calculated and measured limiting collection currents and collection efficiencies for (a) the RRDE system and (b) the DEFC setup with different channel widths of 2.0, 5.0 and 7.5 mm.

**Table 4.** Collection efficiency of the RRDE and DEFC.

| | **Analytical** | **Simulation** | **Experimental fit y = Nx+0** | |
|---|---|---|---|---|
| **Device** | **N [%]** | **N [%]** | **N [%]** | **R2** |
| RRDE | 0.425 | n/a | 0.434(4) | 0.9996 |
| DEFC, w = 2.0 mm | 0.354 | n/a | 0.356(9) | 0.9969 |
| DEFC, w = 5.0 mm | 0.354 | 0.358(1) | 0.344(3) | 0.9996 |
| DEFC, w = 7.5 mm | 0.354 | n/a | 0.351(0) | 1.0000 |

Similar trends were observed for the DEFC with different widths, where the deviation of the collection efficiency was most pronounced for the narrowest width and flow rate < 2.2 mL/min (1.3 mL$^{1/3}$min$^{1/3}$) (Fig. S8b). The dataset taken with w=2.0 mm had the largest experimental spread, which decreased with increasing the width (Fig. S9b-d). Nonetheless, the analytical collection efficiency was found within 1$\sigma$ for this sample, 4$\sigma$ for w=5 mm and 16$\sigma$ for w=7.5 mm using fits to Eq. 10 (Table 4). The simulations of the 5 mm DEFC were in good agreement with the analytical solution but also showed a slight reduction of the collection efficiency with the flow velocity (Fig. S8). This effect is rarely discussed in literature[1] and might be due to the complex velocity-dependent distribution of current density and resultant product concentration above the electrodes (Fig. 3b). The calculated collector current (dashed line) matched the measured collector currents (solid symbols) well for all widths (Fig. 6b). We conclude that all investigated electrode widths can be used for generator-collector experiments where the narrowest electrode is most accurate towards the analytical solution at the cost of precision (i.e., largest experimental error) and the widest electrode the least accurate but most precise. The 5 mm DEFC balances accuracy and precision well. Furthermore, the common analytical solutions of the limiting current and collection efficiency (Eq. 16/18) apply. Therefore, we recommend it as a practical compromise for DEFC experiments.

## Conclusion

We comparatively studied the limiting current and collection efficiency on a commercial RRDE and home-built DEFCs, which produced measurements of similar quality. While comparable results to commercial RRDEs can be achieved, it is important to be aware of the differences in forced convection (rotation speed vs. flow rate) and the non-uniformly accessibility of the DEFC electrodes. The latter means that a non-uniform distribution of active material (e.g. an imperfectly applied catalytic ink) may have a larger effect on the currents in the flow cell as compared to the RRDE system. To circumvent this issue, we herein focused on the ferri-/ferrocyanide redox reaction on Pt electrodes. The measured limiting current of the generator electrode matched that of the analytical solution for the RRDE and DEFCs with 2.0 and 5.0 mm. The 7.5 mm DEFC could not be described by the analytical solution because our data suggested a different exponential dependence on the width due to the aspect ratio of the electrode. We proposed a normalization procedure for the geometry of DEFCs, which can facilitate comparison to past and future DEFC studies with different geometries if the analytical solution applies. The collection efficiencies of the RRDE and DEFC were close to the analytically calculated values of 42.5% and 35.4%, respectively. The 2.0 mm wide electrode showed the most accurate collection efficiency at the cost of precision and the 7.5 mm wide electrode showed the least accurate collection efficiency but was most precise. We conclude that the 5 mm DEFC balanced accuracy and precision best. These dimensions are also well suited for the integration of the DEFC with spectroscopic methods. Alternatively, our design allows for easy removal of the electrodes for post mortem investigations. We conclude that our DEFC with exchangeable electrodes is an attractive alternative to commercial RRDEs due to the flexibility to optimize the electrode materials and geometry for the desired experimental setup and reaction of interest.

## Experimental Section

**Material and chemicals**

The electrolyte consisted of 2 mM potassium hexacyanoferrate from Merck (analytical grade ≥99 %) and 0.1 M potassium chloride from Merck (≥99.5%) in ≥18.2 MΩ Milli-Q® water. The electrolyte was purged with argon (5.0 AirLiquide Alphagaz) for longer than 45 minutes to remove all dissolved oxygen. All chemicals were used as received. During the measurements, the purging was continued by bubbling argon into the reservoir.

**Rotating ring-disk setup**

The experimental setup of the RRDE was composed of a custom-made cylindrical PTFE cell, a RRDE-3A rotator (ALS Japan Co Ltd.), a saturated calomel electrode (RE-2B, ALS Japan Co Ltd.) and a platinum counter electrode, both radially arranged around the RRDE at a distance of 17 mm. The RRDE (ALS Japan Co Ltd.) consisted of a 4 mm diameter platinum disk (A=0.126 cm²) as generating electrode and a concentric platinum ring with an inner diameter of 5 mm and an outer diameter of 7 mm as detection electrode, separated by a PTFE spacer. The disk and ring electrodes were separately polished with $Al_2O_3$ using 3 µm and 0.04 µm polishing slurries and cleaned in isopropanol, before being assembled. During the experiment the rotator was set to various rotation speeds (100, 200, 300, 400 and 500 rpm).

**Flow cell setup**

The cell parts of the DEFC were constructed out of polyoxymethylene (POM) due to its mechanical stability. All cell parts with contact to the electrolyte were cleaned by rinsing first with isopropanol, then with ≥18.2 MΩ Milli-Q® water, subsequently sonicated in ultrapure water for at least five minutes and air dried afterwards. The platinum electrodes (99.95 % purity) had a size of 10x5x1 mm (length, width, thickness). They were polished using 3 µm and 0.04 µm polishing slurry made of $Al_2O_3$, cleaned with ultrapure water and isopropanol by sonication. The electrodes were then placed into the holder and wires were laid on the back of the electrodes for electrical connection. To tighten the system, Parafilm® was placed on top of the wire and electrodes and a PTFE block pressed the assembly together. The glassy carbon counter electrode was cleaned separately the same way as the Pt-electrodes and placed in the holder on top of the working electrodes. The saturated calomel electrode (SCE; RE-2BE ALS Japan Co Ltd.) was placed upstream of the working electrodes. Finally, the assembled cell was purged multiple times with argon to remove the oxygen in the system and the electrolyte was drawn into the channel. The flow in the channel was controlled using a LA100 syringe pump from HLL Landgraf Laborsysteme with a 50 ml Omnifix® syringe from B|Braun. The electrolyte was thereby drawn from a container with a constant argon flow and small overpressure trough the cell into the syringe with flowrates ranging from 0.5 to 6 mL/min. The small overpressure was applied to avoid oxygen from entering the system. To build up and check that the cell was under a small overpressure the container had an extra outlet to an additional container with Milli-Q® water. The overpressure was guaranteed, as long as argon bubbles could be observed in the extra container.

**Electrochemical measurements**

For the electrochemical measurements, two Gamry instrument interfaces 1010 were used in bipotentiostat mode. The potentiostats were calibrated to correct for possible current offsets and cable capacitance. The resistance of the working electrodes was determined using impedance spectroscopy before the measurements and afterwards to detect possible problems with connections and to correct for the potential drop. The generator electrode was then cycled three times between 0.55 V vs. SCE and -0.3 V vs. SCE with a scan rate of 10 mV/s. The collector electrode was held at a constant potential of 0.55 V vs. SCE for the entire measurement. All electrochemical data were analyzed using a custom python script. Impedance correction by iR-subtraction was applied to all measurements where the resistance, R, was measured by impedance spectroscopy. Additionally, a linear baseline of the collector electrode was subtracted from the data to correct for non-faradic and other background currents (Fig. S10 a,b). The generator current was corrected by subtracting the linear function fitted to the data before the onset of the reduction (Fig. S10 c,d).The limiting currents were obtained from the anodic scan for potentials with constant current. For most measurements this was the case between 0 and -0.2 V vs. SCE. In the case of slower flow rates, the potential windows had to be shifted to lower values due to the residue of the reduction peak. However, in no case was a limit lower than -0.28 V vs. SCE chosen to avoid possible artefacts from changing the scan direction.

**Simulations**

3D fluid dynamic simulations were first performed based on the geometry of our 5 mm experimental DEFC (further parameters in Table 1). Because our flow cell has a sufficient entrance length to develop laminar flow before reaching the electrodes (Fig. S1), constant velocity field shown in Fig. 3a was applied over the electrodes in the following mass transport simulation. With the simulated velocity field, the mass transport equation (Eq. 4) was numerically solved in 3D models with the limiting boundary conditions of $[Fe(CN)_6^{3-}] = 0$ on the generator and $[Fe(CN)_6^{4-}] = 0$ on the collector electrode. For a fair comparison between the limiting currents in 2D and 3D simulations, comparable mesh sizes were introduced. The mesh size close to the electrode was obtained from a previous report[1] as shown in Fig. S11. This resulted in a large number of meshes in the 3D model; the channel was then partitioned into a series of boxes of 1 mm length, in order to make the calculations more manageable (Fig. S11). The concentration profile at the outlet of the previous box was used as the boundary condition at the inlet of the succeeding box. We validated this approach in a 2D model (i.e. without walls), where the same limiting currents and collection efficiency were obtained for the separated channel and single full channel (results are not shown). Local current density is obtained from the flux at the electrode by Eq. 2. The simulated limiting current densities on the generator and the collector were further introduced to 3D model to determine the local current density on the counter electrode, which then determines the flux of products from the counter electrode.

$$\nabla \cdot \vec{\boldsymbol{J_l}} = 0 \qquad (23)$$

$$\vec{\boldsymbol{J_l}} = -\sigma \nabla \phi_l \qquad (24)$$

where $\vec{\boldsymbol{j}}_l$, $\sigma$ and $\phi_l$ represent the ionic current vector, electrolyte conductivity, and electrolyte potential, respectively. Local current density at the electrode surface ($j_s$) is described by

$$\vec{\boldsymbol{n}} \cdot \vec{\boldsymbol{J_l}} = j_s \qquad (25)$$

where $\vec{\boldsymbol{n}}$ denotes the normal vector to the boundary. The electrode was assumed to be highly conductive (i.e., no Ohmic loss). The potential of the counter electrode ($\phi_s$) was set to 0 V. At the counter electrode surface, the following Butler-Volmer equation was applied

$$j_s = j_0 \left\{ exp\left(\frac{\alpha_a F \eta}{RT}\right) - exp\left(\frac{-\alpha_c F \eta}{RT}\right) \right\}, \qquad (26)$$

in which $j_0$, $\alpha_a$, and $\alpha_c$ are the exchange current density, anodic and cathodic transfer coefficient, respectively. The overpotential ($\eta$) is described as follow.

$$\eta = \phi_s - \phi_l - E_{eq} \qquad (27)$$

$E_{eq}$ is the equilibrium potential. Since the reduction current on the generator is larger than the oxidation current on the collector, an additional anodic reaction takes place on the counter electrode to account for charge balance. This reaction is assumed to be the oxygen evolution reaction. The simulated local current density on the counter electrode determines the boundary condition for the flux of $O_2$.

$$\frac{j_s}{nF} = \vec{\boldsymbol{n}} \cdot \vec{\boldsymbol{J}} \qquad (28)$$

where $\vec{\boldsymbol{J}}$ is the flux vector. For simplicity, we ignored any concentration overpotentials due to pH gradient and assumed $O_2$ remains in the electrolyte as dissolved gases (i.e., no bubbles). We confirmed that these simplifications do not affect the limiting currents at the generator and collector, which are the main focus of this study.

## Acknowledgements

This work received funding from the Deutsche Forschungsgemeinschaft (DFG, German Research Foundation) – 397636017 and under Germany´s Excellence Strategy – EXC 2008/1 (UniSysCat) – 390540038 as well as from the German Helmholtz Association – Excellence Network – ExNet-0024-Phase2-3. The simulation work was also carried out with the support of the Helmholtz Energy Materials Foundry (HEMF), a large-scale distributed research infrastructure founded by the German Helmholtz Association. We like to thank Prof. Gunther Wittstock for helpful discussions.

**Keywords:** analytic electrochemistry • flow cells • rotating-ring disk electrodes • Multiphysics simulations • in situ experiments

**Supporting information**

# Characterization of a modular flow cell system for electrocatalytic experiments and comparison to a commercial RRDE system

Frederik J. Stender[a], Keisuke Obata[b], Max Baumung[a], Fatwa F. Abdi[b], Marcel Risch[a,c]*

[a] Frederik Johannes Stender, Max Baumung, Dr. Marcel Risch
Institut für Material Physik, Georg-August-Universität Göttingen, Friedrich-Hund-Platz 1, 37085 Göttingen
[b] Dr. Keisuke Obata, Dr. Fatwa Firdaus Abdi
Institut für Solare Brennstoffe, Helmholtz-Zentrum Berlin für Materialien und Energie GmbH, Hahn-Meitner-Platz 1, 14109 Berlin
[c] Dr. Marcel Risch, Nachwuchsgruppe Gestaltung des Sauerstoffentwicklungsmechanismus, Helmholtz-Zentrum Berlin für Materialien und Energie GmbH, Hahn-Meitner-Platz 1, 14109 Berlin
E-mail: marcel.risch@helmholtz-berlin.de

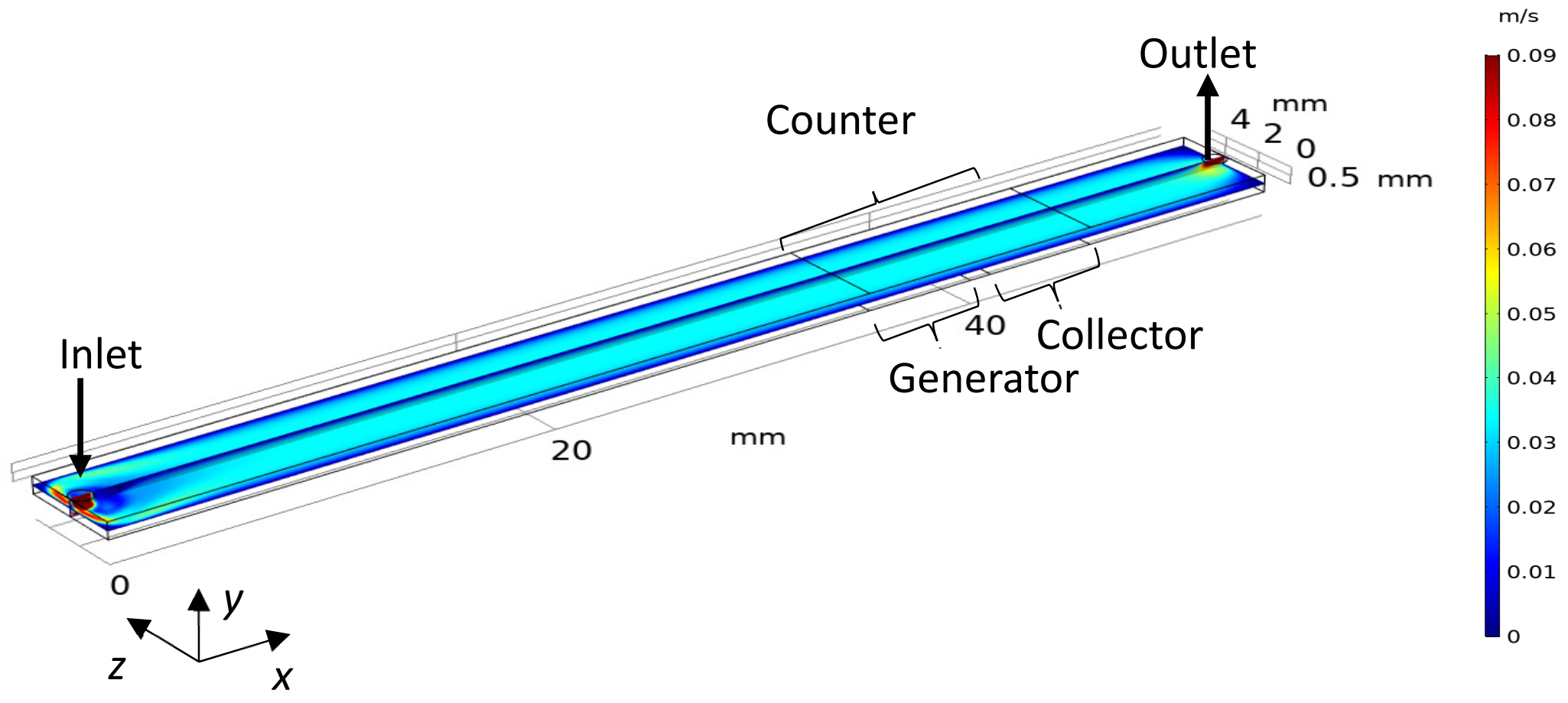

**Figure S1**. 3D simulations of the velocity distribution in the 5 mm DEFC channel.

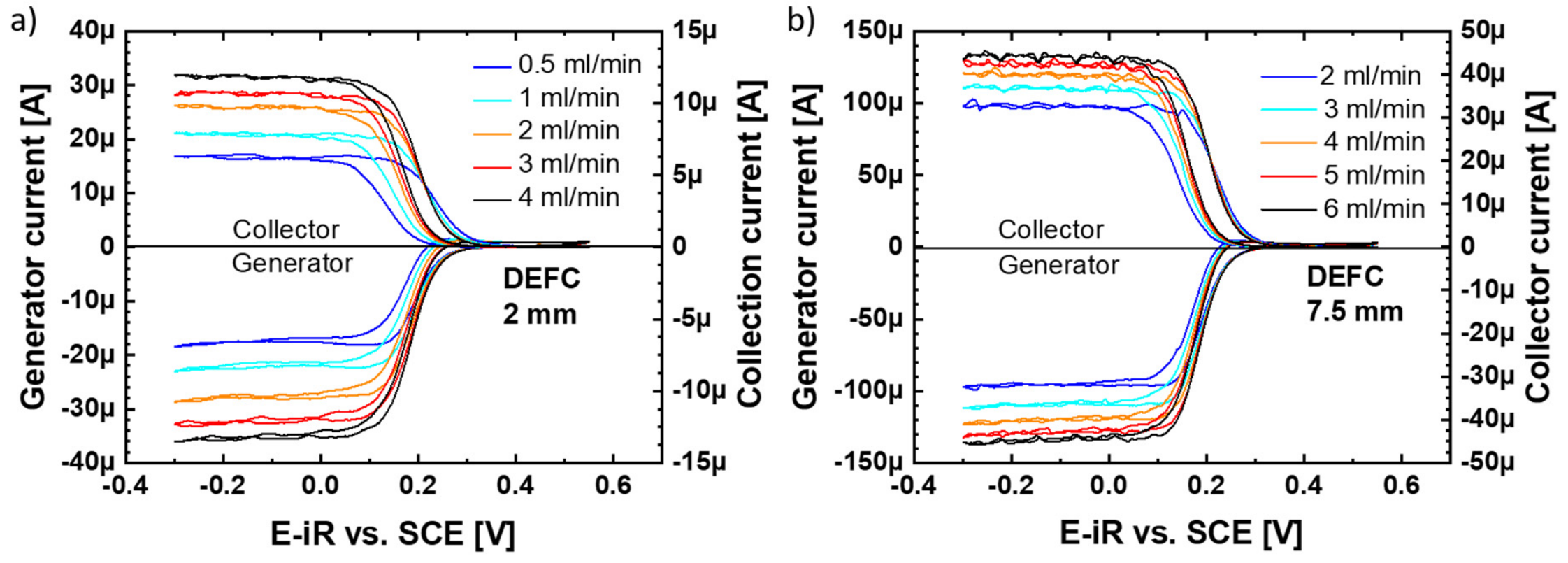

**Figure S2.** Cyclic voltammograms of the 2 mm and 7.5 mm DEFC for different flow rates. The scan range was between 0.55 and -0.3 V vs SCE.

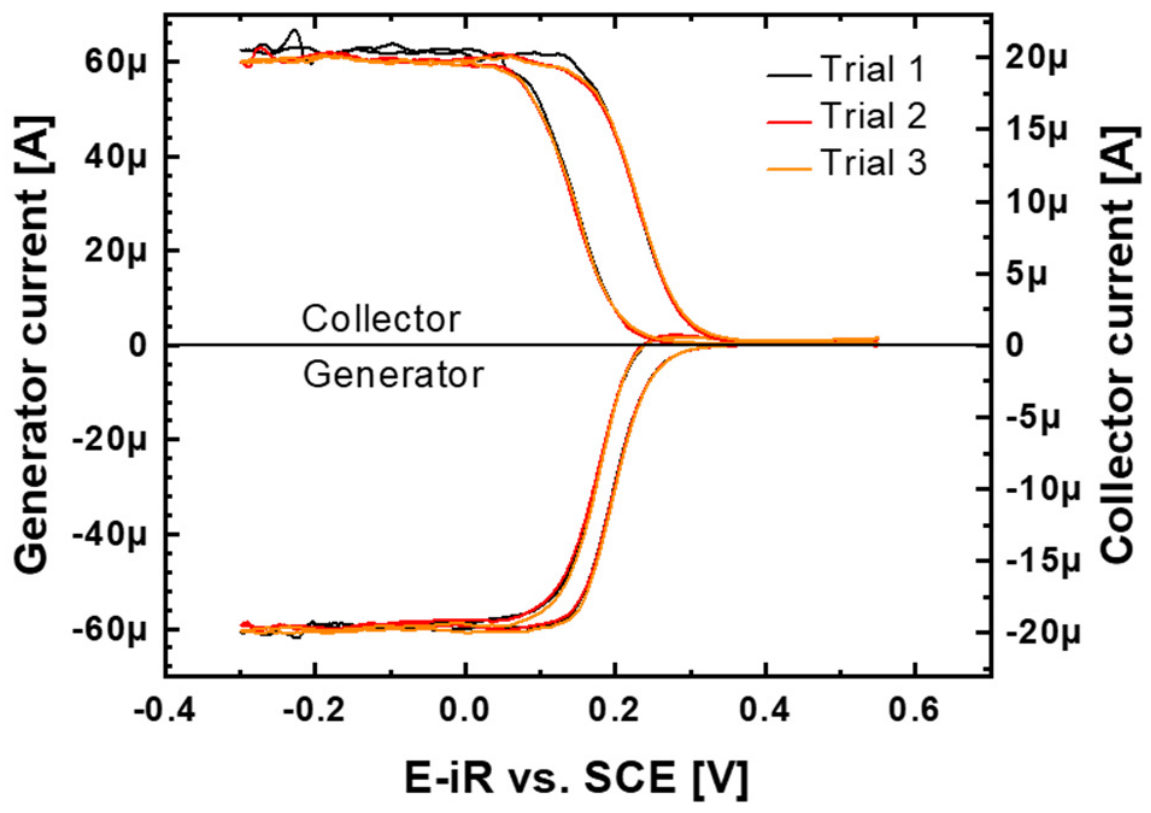

**Figure S3.** Reproducibility test for the 5 mm DEFC. The trials were all independently assembled setups with polished electrodes and fresh electrolyte measured on different lab days.

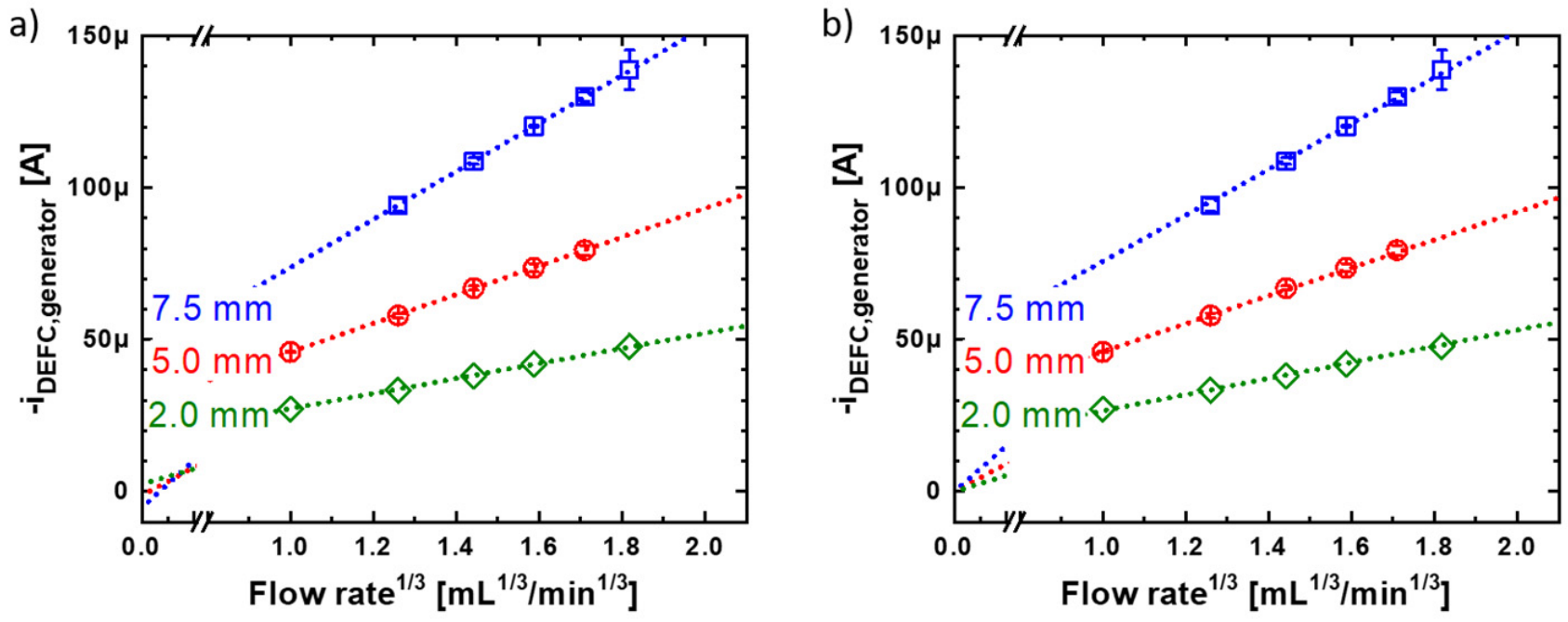

**Figure S4.** Linear fits of the limiting current with (a) a linear fit including a finite y-axis intercept and (b) with a fit through the origin for all three DEFC channel widths.

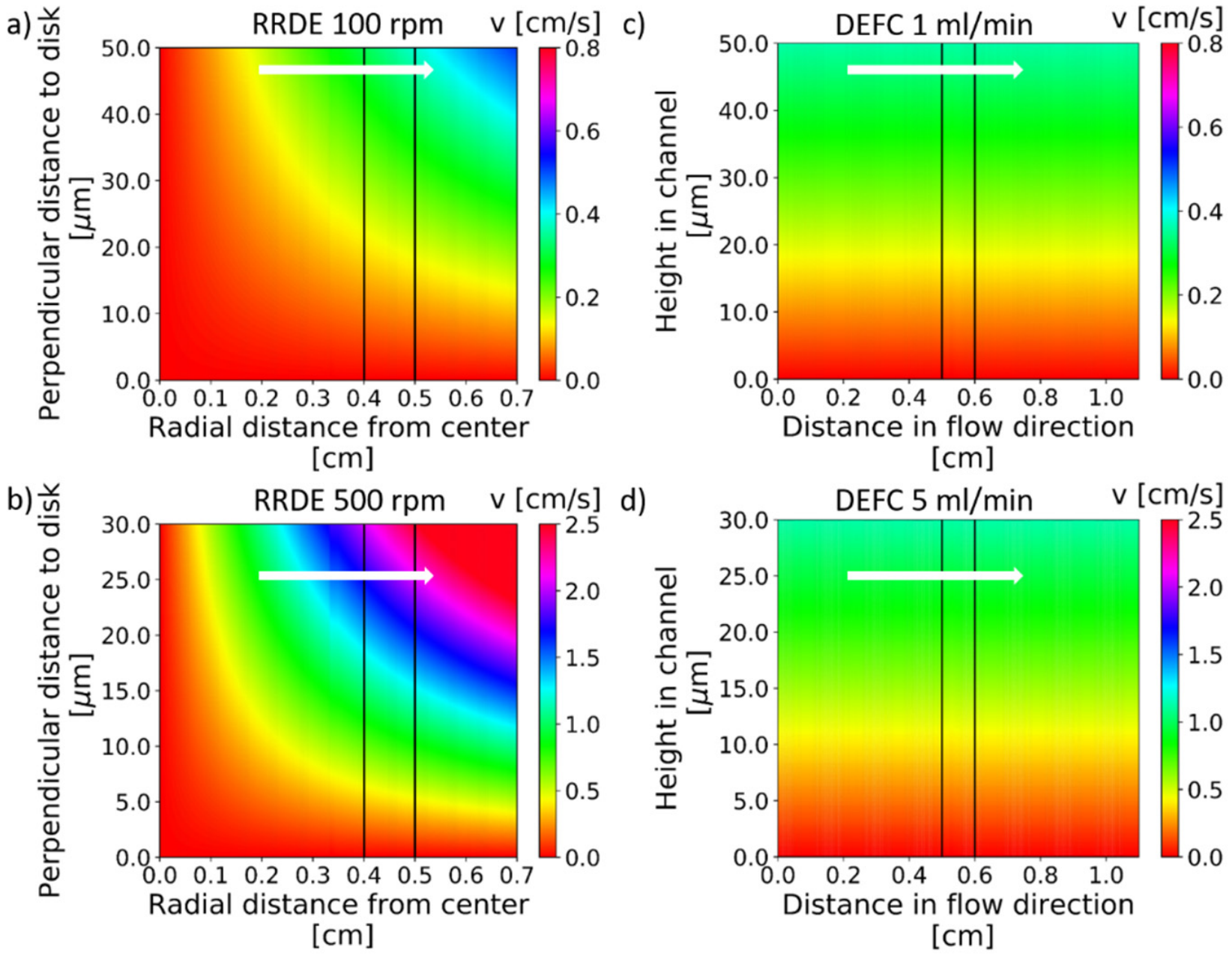

**Figure S5.** Analytically calculated velocity profiles using Eqs. 6 and 14 for the RRDE in radial direction at (a) 100 rpm and (b) 500 rpm as well as for the 5 mm DEFC in the middle of the channel (z=d/2) in x-direction for (c) 1 mL/min and (d) 5 mL/min. The DEFC images are shown from the beginning of the generator electrode (x=0) to the end of the collector electrode (x=1.1). The RRDE images are shown from the center of the generator electrode (disk; r =0.0 cm) and to the outer rim of the collector electrode (ring; x=0.7 cm). The black lines mark the end of the generator electrode and the beginning of the collector electrode. The white arrow indicates the flow direction.

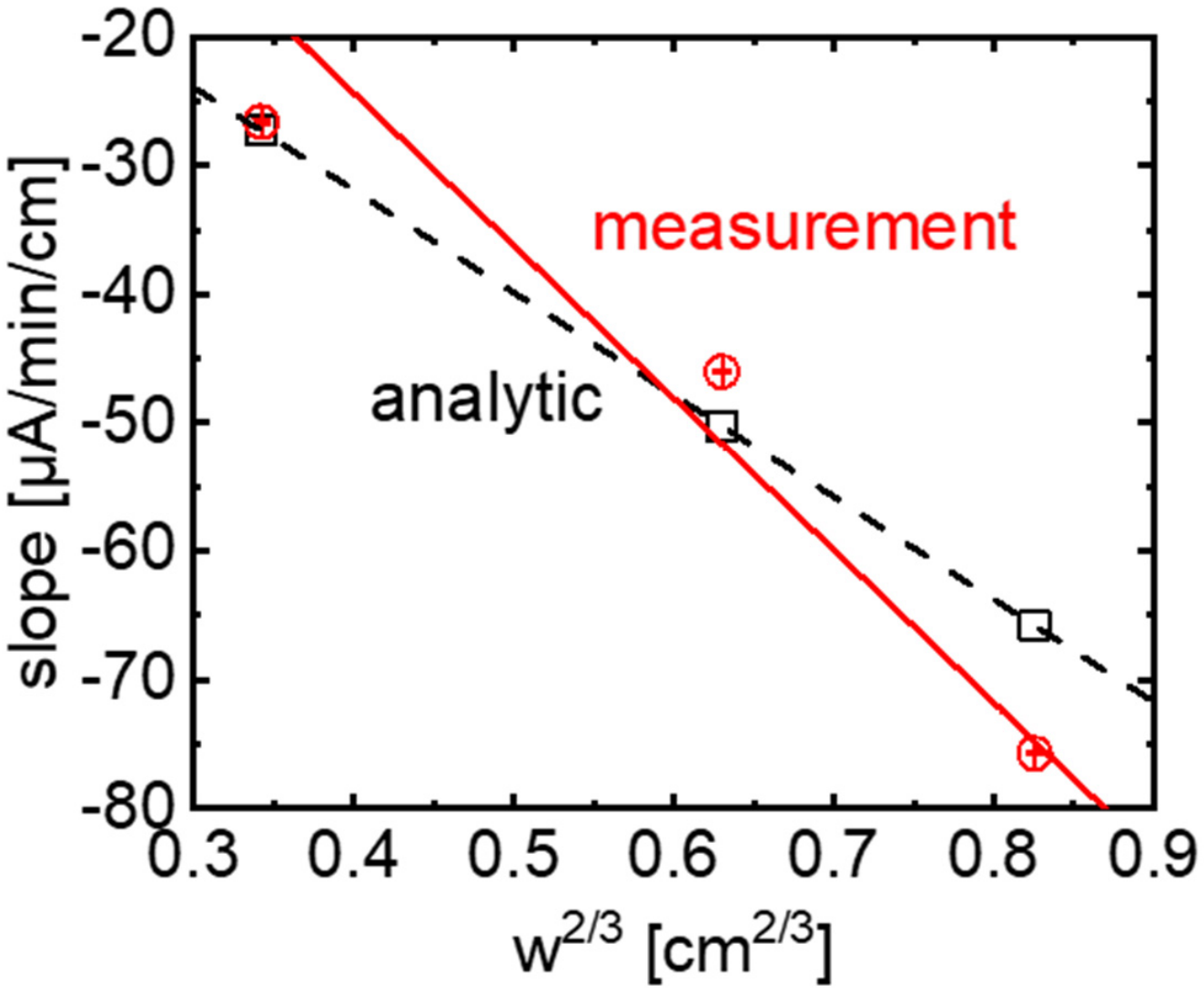

**Figure S6.** Analytical and measured slopes of the limiting currents (Table 3) plotted over the expected width of channel behavior (Eq. 20)

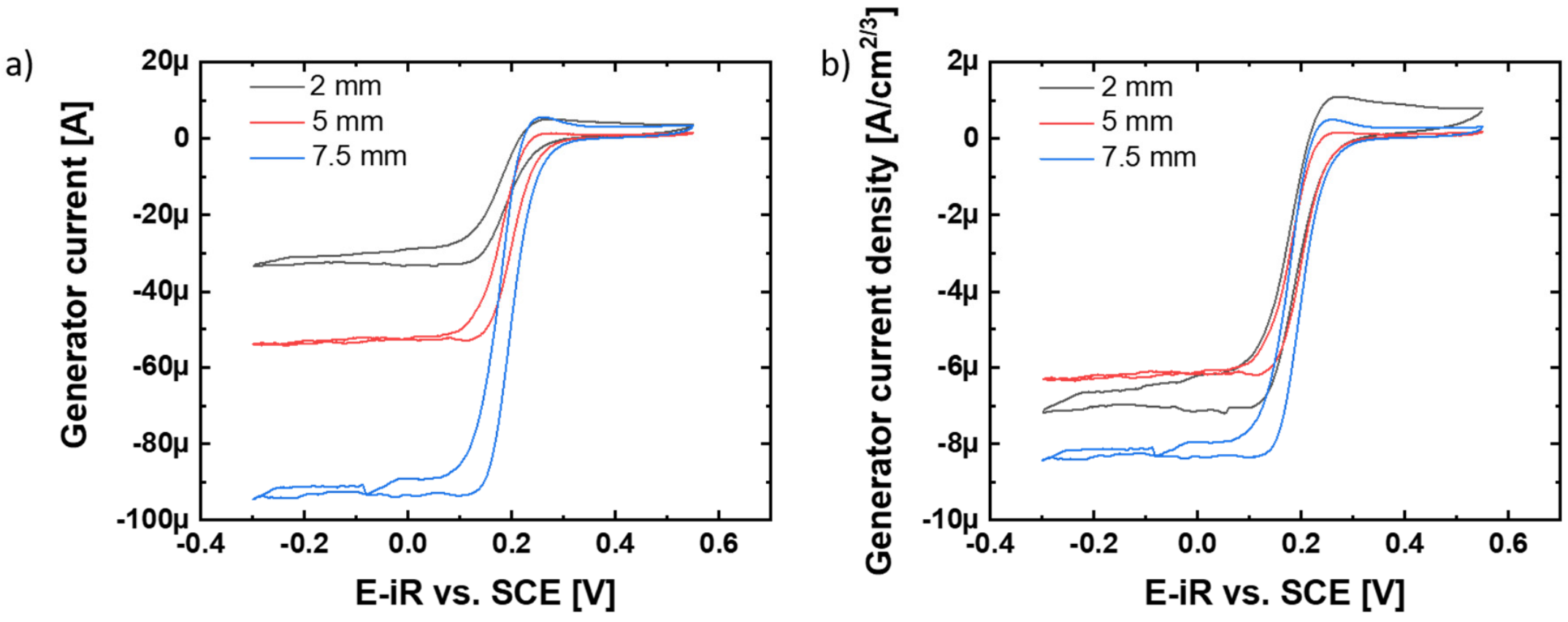

**Figure S7.** (a) Non-normalized and (b) normalized measurements of different channel widths of the DEFC for 2 mL/min. The normalization was done according to Eq. 22.

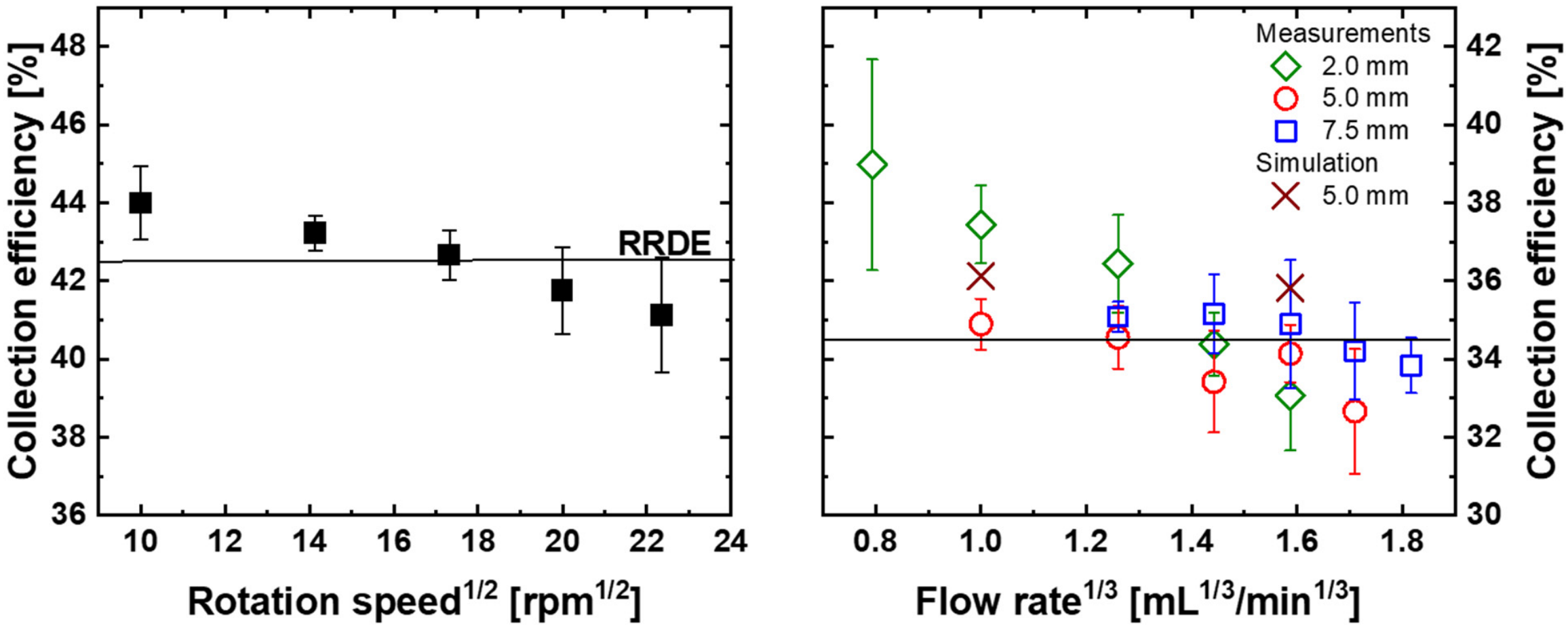

**Figure S8.** Calculated and measured collection efficiencies (a) for the RRDE at various rotation speeds and (b) for the DEFCs at various flow rates. Solid line indicates the value estimated from the analytical solution.

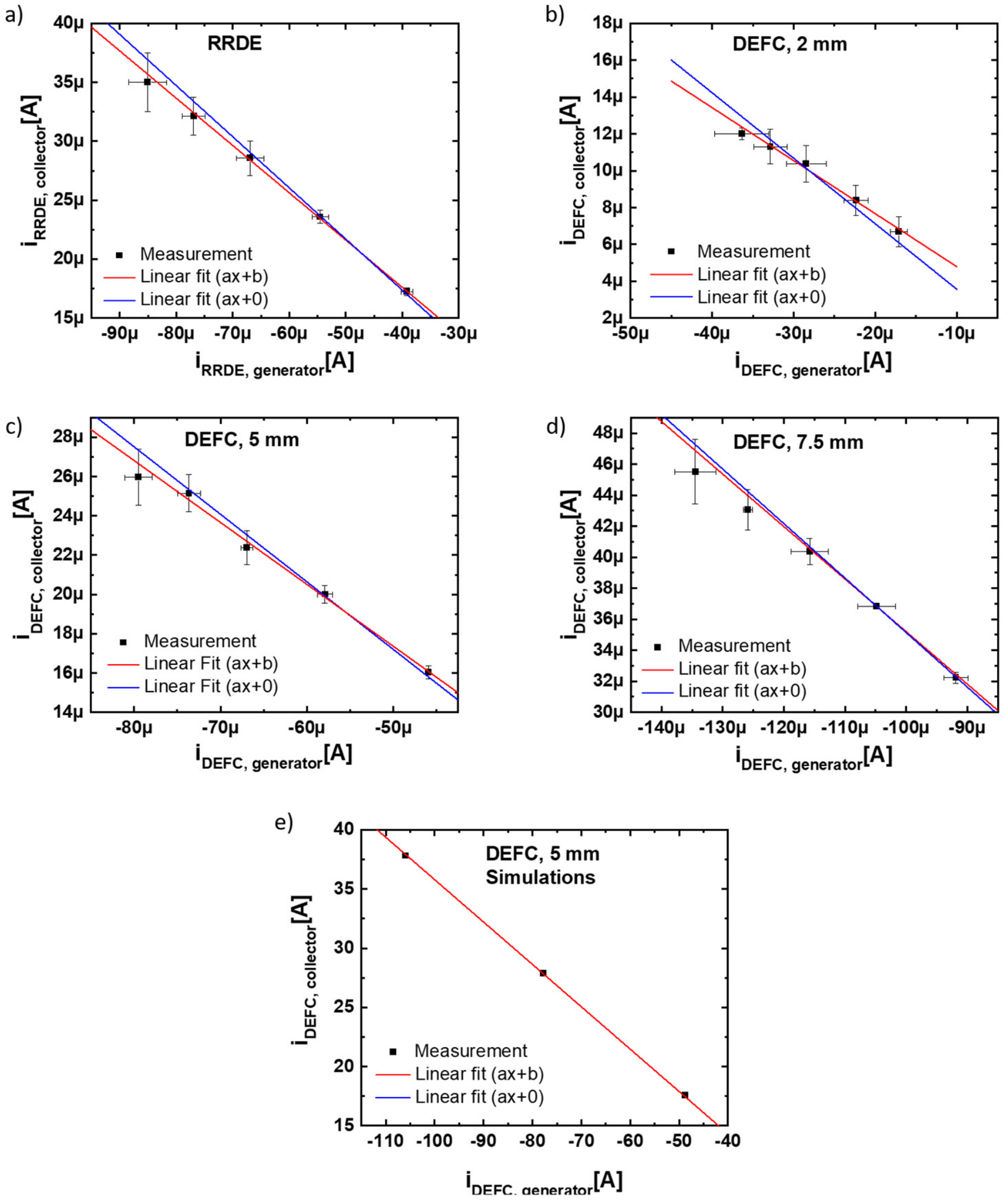

**Figure S9.** Linear fits with and without a finite y-axis intersection for (a) the RRDE setup and DEFC setups measurements with widths of (b) 2.0 mm, (c) 5.0 mm, (d) 7.5 mm as well as (e) the simulation of the 5 mm DEFC.

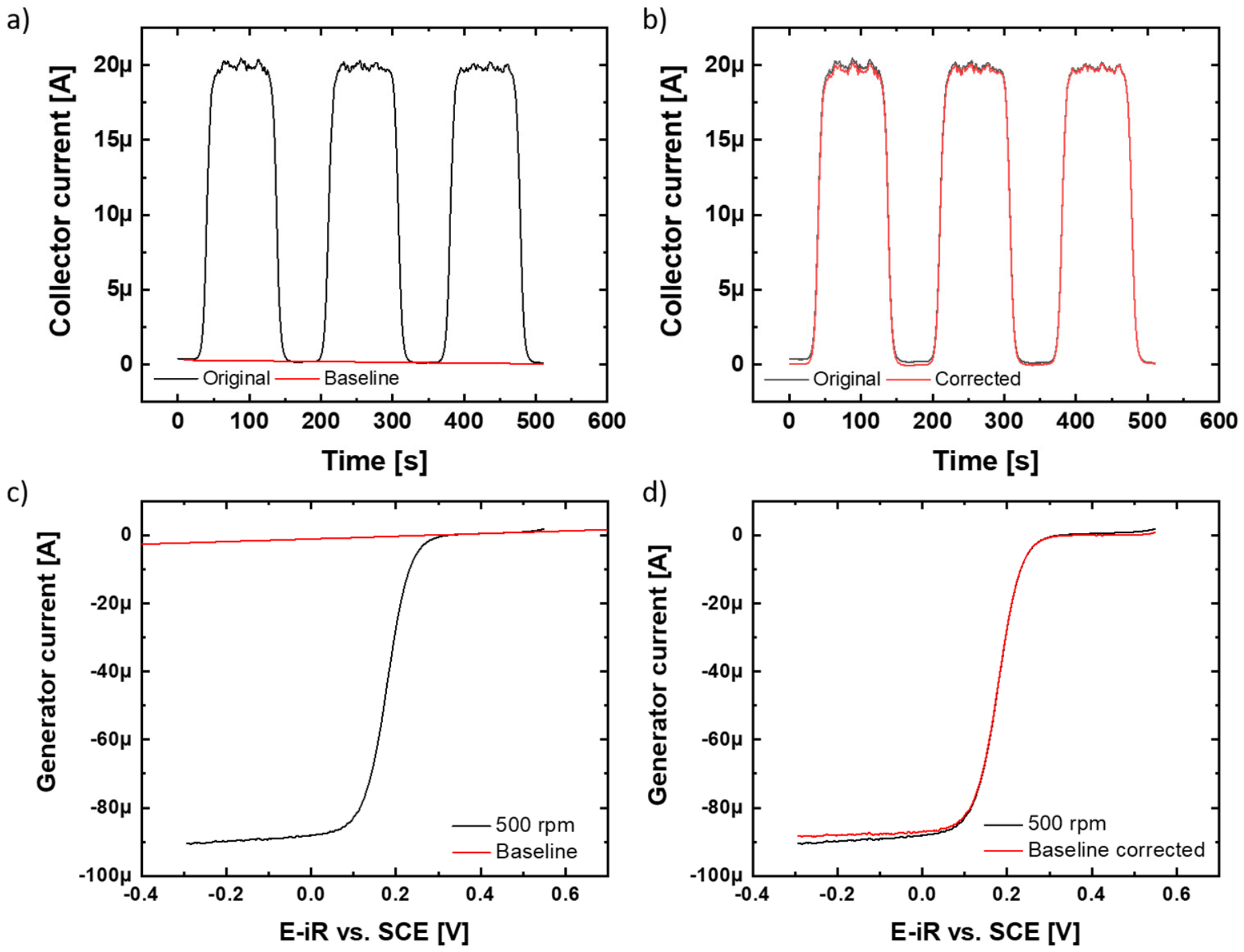

**Figure S10** Background correction to correct for non-faradic currents at the generator and collector electrode. (a,c) First a baseline is fitted to the data before the oxidation/reduction and then subtracted from the original data. (b,d) The corrected data in comparison to the original data.

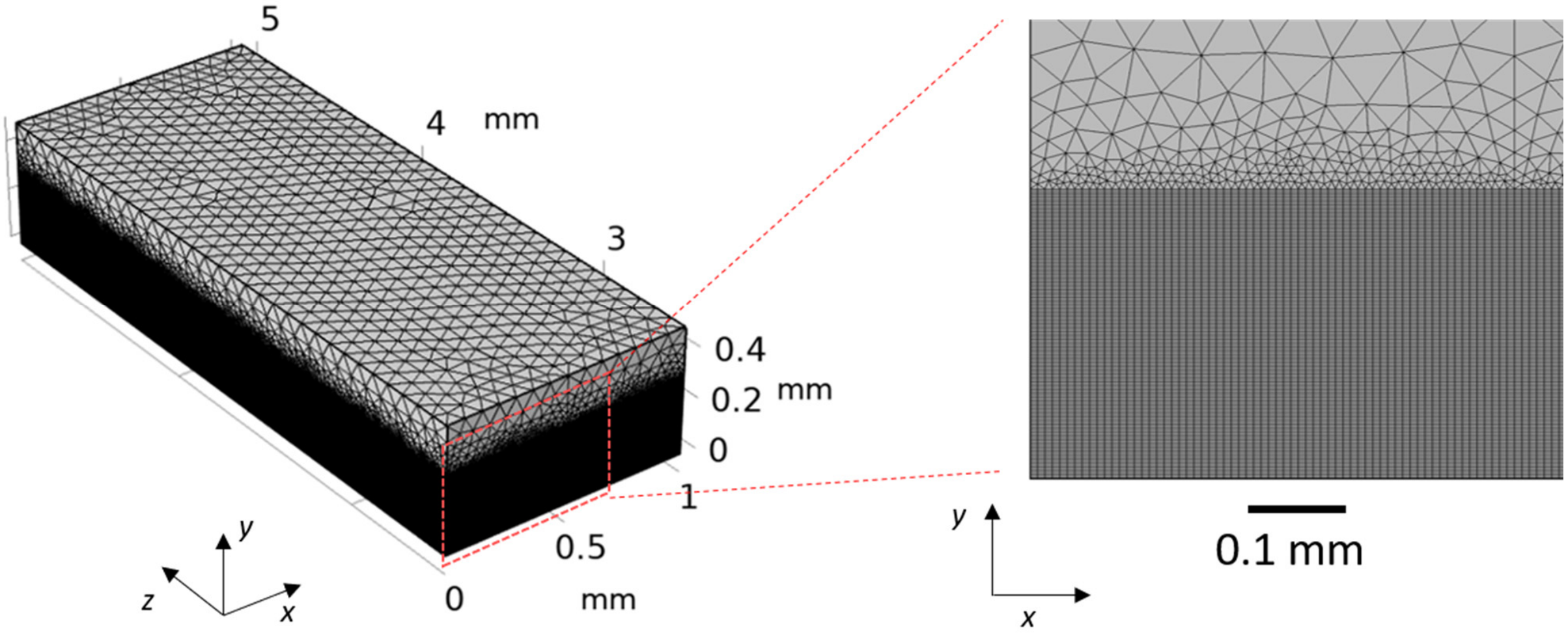

**Figure S11.** The 1 mm mesh block that was used for the 3D simulations.

**Table S1.** Comparison between the experimental, calculated and simulated fits of the limiting currents in Fig. S4 with and without finite y-axis intercept.

| | **Analytic*** | **Simulation** | **Experimental fit 1**<br>**y = ax+0** | | **Experimental fit 2**<br>**y = ax+b** | | |
|---|---|---|---|---|---|---|---|
| **Device** | **slope** | **slope** | **Slope a** | **$R^2$** | **slope a** | **Y intercept b** | **$R^2$** |
| RRDE | -3.96 μA/rpm$^{1/2}$ | n/a | -3.86(2) μA/rpm$^{1/2}$ | 0.9999 | -3.76(3) μA/rpm$^{1/2}$ | -1.54(37) μA | 0.9998 |
| DEFC, w = 2.0 mm | -27.3 μA*min$^{1/3}$/cm | n/a | -26.6(2) μA*min$^{1/3}$/cm | 0.9996 | -24.8(4) μA*min$^{1/3}$/cm | -2.48(55) μA | 0.9987 |
| DEFC, w = 5.0 mm | -50.2 μA*min$^{1/3}$/cm | -49.5 μA*min$^{1/3}$/cm | -46.0(1) μA*min$^{1/3}$/cm | 1.0000 | -47.3(3) μA*min$^{1/3}$/cm | 1.38(28) μA | 1.0000 |
| DEFC, w = 7.5 mm | -65.8 μA*min$^{1/3}$/cm | n/a | -75.7(0) μA*min$^{1/3}$/cm | 1.0000 | -79.1(2) μA*min$^{1/3}$/cm | 5.37(26) μA | 1.0000 |

**Table S2.** Collection efficiencies for the calculated, simulated and measured RRDE and DEFC systems in Fig. 9 determined from linear fits with and without finite y-axis intercept.

| | **Analytic*** | **Simulation** | **Experimental fit 1**<br>**y = Nx+0** | | **Experimental fit 2**<br>**y = Nx+b** | | |
|---|---|---|---|---|---|---|---|
| **Device** | **N** | **N** | **N** | **$R^2$** | **N** | **Y intercept b** | **$R^2$** |
| RRDE | 0.425 | n/a | 0.434(4) | 0.99959 | 0.401(6) | 1.55 (30) E-6 | 0.99898 |
| DEFC, w = 2.0 mm | 0.354 | n/a | 0.356(9) | 0.99688 | 0.288(17) | 1.90(48) E-6 | 0.98543 |
| DEFC, w = 5.0 mm | 0.354 | 0.358(1) | 0.344(3) | 0.99958 | 0.315(11) | 1.60(62) E-6 | 0.99481 |
| DEFC, w = 7.5 mm | 0.354 | n/a | 0.351(0) | 1.0000 | 0.339(15) | 1.34 (157) E-6 | 0.99226 |